\documentclass[final,3p,times,twocolumn]{elsarticle}

\usepackage{amssymb}

\journal{Nuclear Instruments and Methods in Physics Research A}

\begin{document}

\begin{frontmatter}


 \title{FEL performance and tolerance studies of the\\
EuPRAXIA@SPARC\_LAB beamline AQUA}
 \author[aff1,aff2]{Federico Nguyen\corref{cor1}}
 \ead{federico.nguyen@enea.it}
 \cortext[cor1]{Corresponding Author}
 \affiliation[aff1]{organization={ENEA},
             addressline={Via Enrico Fermi 45},
             city={Frascati},
             postcode={00044},
             country={Italy}}
 \affiliation[aff2]{organization={LNF-INFN},
             addressline={Via Enrico Fermi 54},
             city={Frascati},
             postcode={00044},
             country={Italy}}

\author[aff2,aff3]{Luca Giannessi}
\affiliation[aff3]{organization={Elettra--Sincrotrone S.C.p.A.},
            addressline={Strada Statale 14 -- km 163,5}, 
            city={Basovizza},
            postcode={34149}, 
            country={Italy}}
\author[aff2]{Michele Opromolla}
\author[aff1,aff2]{Alberto Petralia}

\begin{abstract}
The AQUA beamline of the EuPRAXIA@SPARC\_LAB facility
is a SASE free-electron laser designed to operate in the water window,
in the 3-4 nm wavelength range. The electron beam driving this source is accelerated up to about 1-1.2 GeV by an X-band
normal conducting linear accelerator, followed by a plasma wakefield acceleration stage.
The main radiator consists of an array of ten APPLE-X permanent magnet
undulator modules, each 2 m long and with a period length of 18 mm.
Tolerance analyses against resistive wall wakefields and injection misalignments
at undulator entrance are performed, and the related effects on the laser yield
performance are evaluated and discussed.
\end{abstract}




\end{frontmatter}


\section{Introduction: design of the AQUA beamline}
\label{sec:1}
The EuPRAXIA project is expected to provide the first Free-Electron Laser (FEL) facility
based on the high accelerating plasma wakefield gradient~\cite{Assmann:2020smm}.
The EuPRAXIA@SPARC\_LAB facility will
be constructed at the INFN-LNF laboratory~\cite{Ferrario:2018gra} and
is going to exploit an electron driver beam to create the wakefield, which then accelerates a following bunch of electrons
to high energies over very short distance. 
The AQUA FEL line is designed to achieve self-amplified spontaneous emission (SASE) to reach for the carbon K-edge,
and to probe the so called water window spectral range, around 3-4 nm wavelength, \textit{i.e.} 310-410 eV photon energy.
Furthermore, the chance to produce selectable polarization radiation allows to study~\cite{Roussel:2017}
chemical properties of materials by means of switchable FEL polarization.
Therefore, the undulator technology chosen for AQUA is a variable polarization permanent magnet APPLE-X,
of the Advanced Planar Polarized Light Emitter (APPLE) type. 
Preliminary calculations, based on the average electron beam parameters, indicate that a period length of 18 mm enables efficient photon flux exploration within the water window, while also offering selectable polarization and some contingency in the total active length.

The design under consideration~\cite{fel2022-wep38} envisages a 2 m long module, namely about 110 periods,
with variable polarization and deflection strength parameter $K$ that can be accordingly tuned to a maximum value:
$K_{max}=1.2$ or $K_{max}=1.7$ in case of circular (CP) or linear (LP) polarization.
The undulator structure is made of four Neodymium-Iron-Boron (NdFeB) permanent magnet blocks with 
a remanent field $B_r=1.35$ T.
The magnets are disposed radially at equal distance around the electron beam axis.
The resulting square hole at the center of the structure allows the installation of a cylindrical vacuum pipe for the propagation of the electrons,
whose diameter is chosen to mitigate the wakefield effects (see Section~\ref{sec:3}).
This configuration enables wavelength tuning properties comparable to other APPLE-type designs with even higher remanent field strength,
but different gap geometry structure~\cite{XLSD5.1}.
\begin{figure}[!htb]
\centering
\includegraphics*[width=1.05\columnwidth]{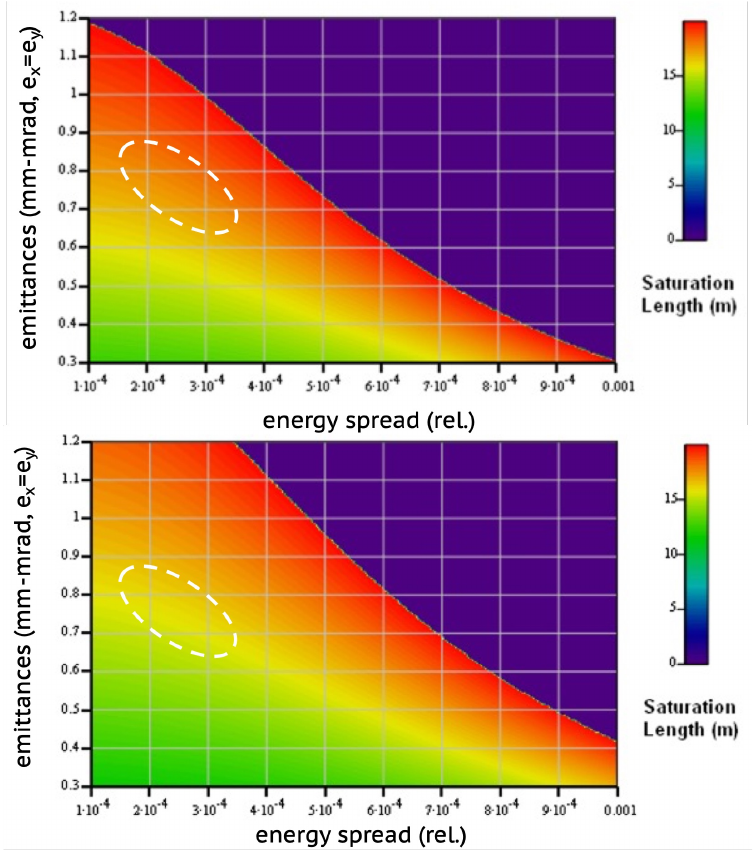}
\caption{FEL saturation length as a function of the fractional energy spread and normalized transverse emittance, for both linear (top) and
circular (down) polarization undulator configurations. The white dashed ellipse indicates the operation region of the AQUA beamline.}
\label{fig:1}
\end{figure}

In order to assess the AQUA FEL performance as a function of the electron beam parameters at undulator entrance, 
the saturation length 
is analyzed with the semi-analytical~\footnote{Xie formulae are modified for a correction factor~\cite{Giannessi:PCTbS} tailored for large beam energy spread cases. This correction factor introduces only a minor deviation from the original Xie result.} formulae~\cite{Dattoli:1984,Xie:1995kp,Dattoli:2004,Giannessi:PCTbS}.
Given the essential role of emittance -- assumed equal in transverse horizontal and vertical planes -- and energy spread,
the resonant wavelength $\lambda=4$ nm working point -- performed with $E_{beam}=1$ GeV electron beam energy,
$I_{peak}=1.5$ kA peak current and average Twiss coefficients $\beta_x=\beta_y=10$ m -- is studied 
by scanning normalized transverse emittance and relative energy spread values,
and assuming Gaussian distributions in current, energy, transverse momenta profiles.

Figure~\ref{fig:1} shows the FEL performance in terms of the saturation length as a result of the parameter scan in energy spread
and transverse emittance, upon setting the undulator beamline in either LP (top panel) or CP (down panel) configurations.
For the expected values of normalized emittance around 0.6-0.9 mm-mrad
and fractional energy spread around $1.5\times10^{-4}$-$3.5\times10^{-4}$, the
saturation length results in 25-28 m or 15-20 m, respectively for LP or CP operations.
The region of expected emittance and energy spread values is highlighted by the white dashed line ellipse.
This length implies that the AQUA design consists of ten APPLE-X modules.

\section{Resistive wall wakefield effects on AQUA}
\label{sec:3}
The vacuum chamber (VC) design is constrained by the resistive wall (RW) wakefields, whose detrimental effects depend on the VC inner radius.
The longitudinal wakefield causes an increase in energy spread that is independent of the beam orbit,
while the transverse wakefield generates a kick angle that depends on the bunch trajectory. 
\begin{figure}[!htb]
\centering
\includegraphics*[width=0.83\columnwidth]{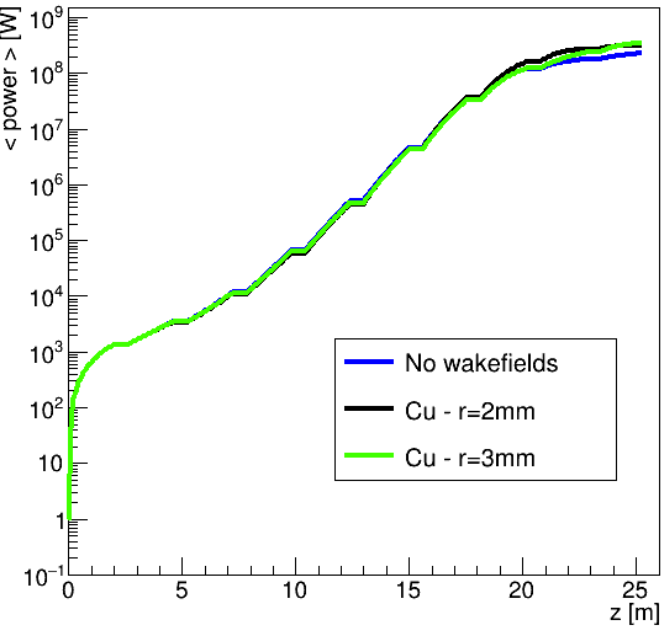}
\caption{Average FEL power growth for VC inner radius of 2 mm (black) and 3 mm (green), compared with the no wakefield case (blue).}
\label{fig:2}
\end{figure}
Assuming a copper VC with cylindrical symmetry and an electron Gaussian current profile with bunch charge $Q=30$ pC and RMS length $\sigma_z=2\,\mu$m,
the energy spread induced by the RW longitudinal wakefield does not affect the FEL power growth along the propagation coordinate.
Figure~\ref{fig:2} shows the average power growth as evaluated from 3D time
dependent simulations with the \texttt{Genesis1.3} code~\cite{Reiche1999}.
The curves obtained by evaluating RW wakefields~\cite{Bosco:2023svr} for both 2 mm and 3 mm inner radius options are
superimposed to the case with no wakefields.
\begin{figure}[!htb]
\centering
\includegraphics*[width=0.95\columnwidth]{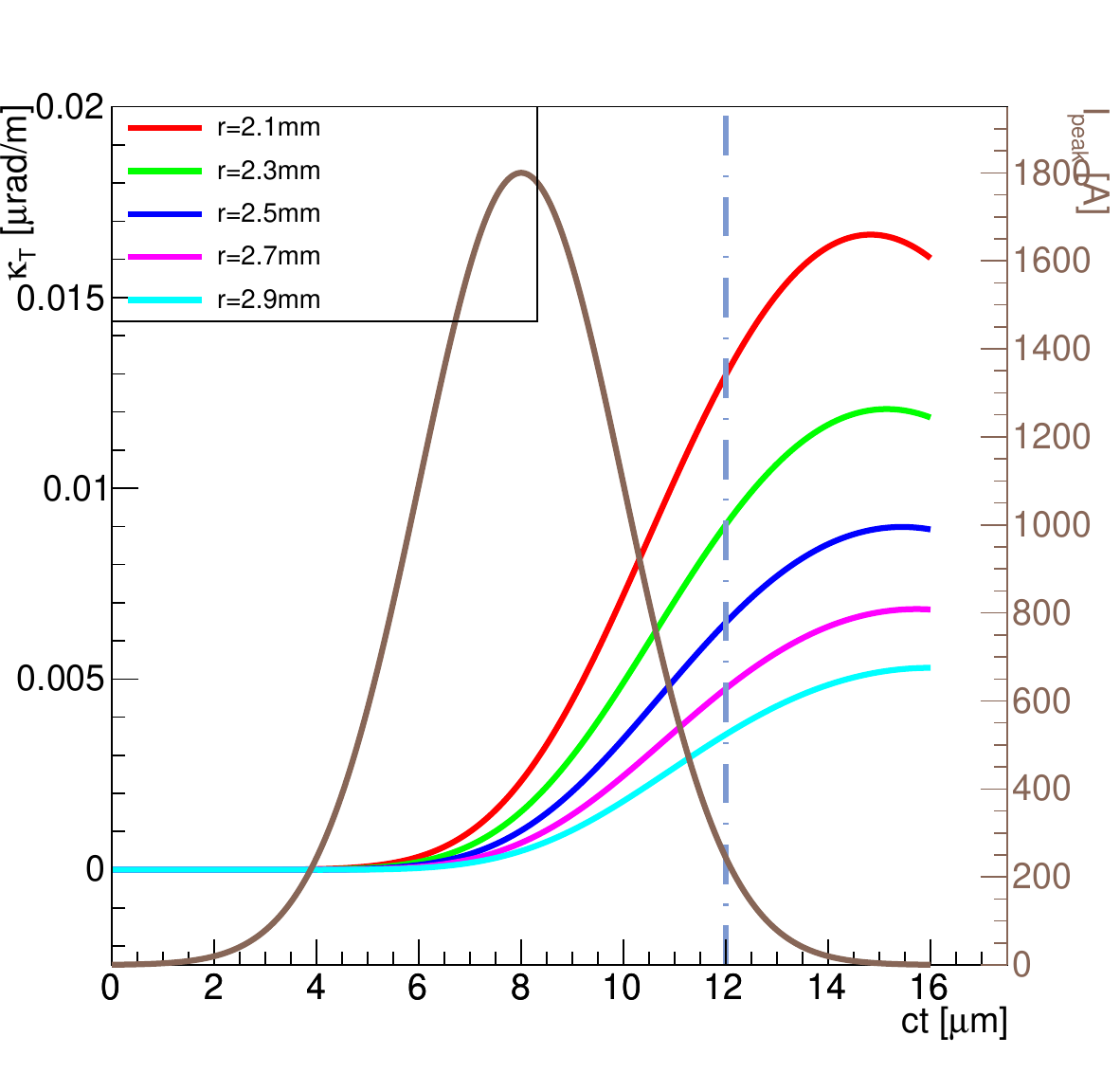}
\caption{Kick angle per unit length for different vacuum chamber inner radii, and for a transverse offset of 50 $\mu$m, superimposed to the short bunch current time profile (brown curve). The vertical dot-dashed line specifies the $2\sigma_z$ coordinate away from the peak position.}
\label{fig:3}
\end{figure}

Within the short-range approximation~\cite{Bane:2004nb},
analytical formulae are obtained by exploiting the relationship~\cite{Bane:2006ce} between the RW longitudinal 
$W_\parallel$ and transverse $W_\perp$ wakefields, inside a vacuum pipe of inner radius $r$:
\begin{equation}
\label{eq:RW1}
W_\perp (s,r) = \frac{2}{r^2}\int_0^s W_\parallel (s^\prime,r) ds^\prime
\end{equation}
where $s$ indicates the intra-bunch coordinate.
Eq.(\ref{eq:RW1}) allows to quantify the impact of transverse wakefields in terms of the kick angle per unit length $\kappa_T\,[rad/m]$,
for a given transverse $\ell_{\rm off}$ offset:
\begin{equation}
\label{eq:RW2}
\kappa_T(s,r) = \ell_{\rm off}\,\frac{e Q}{E_{beam}}\int_{-\infty}^s W_\perp(s-s^\prime,r)\,\rho(s^\prime) ds^\prime
\end{equation}
where $\rho(s)$ is the line charge density and the dependence on the inner radius is included in the wakefield function. 
Figure~\ref{fig:3} shows the current profile as well as the kick angle per unit length, $\kappa_T$, as a function of the bunch time, for different copper VC inner radii, and for a transverse offset of 50 $\mu$m.
For every considered $r$ value, $\kappa_T$ begins to increase well behind the peak of the bunch current.
The line at $2\sigma$ in the current profile of Figure~\ref{fig:3} marks the point along the intra-bunch coordinate where Eq.(\ref{eq:RW2}) is used
\begin{figure}[!htb]
\vspace*{-2.5mm}
\centering
\includegraphics*[width=1.1\columnwidth]{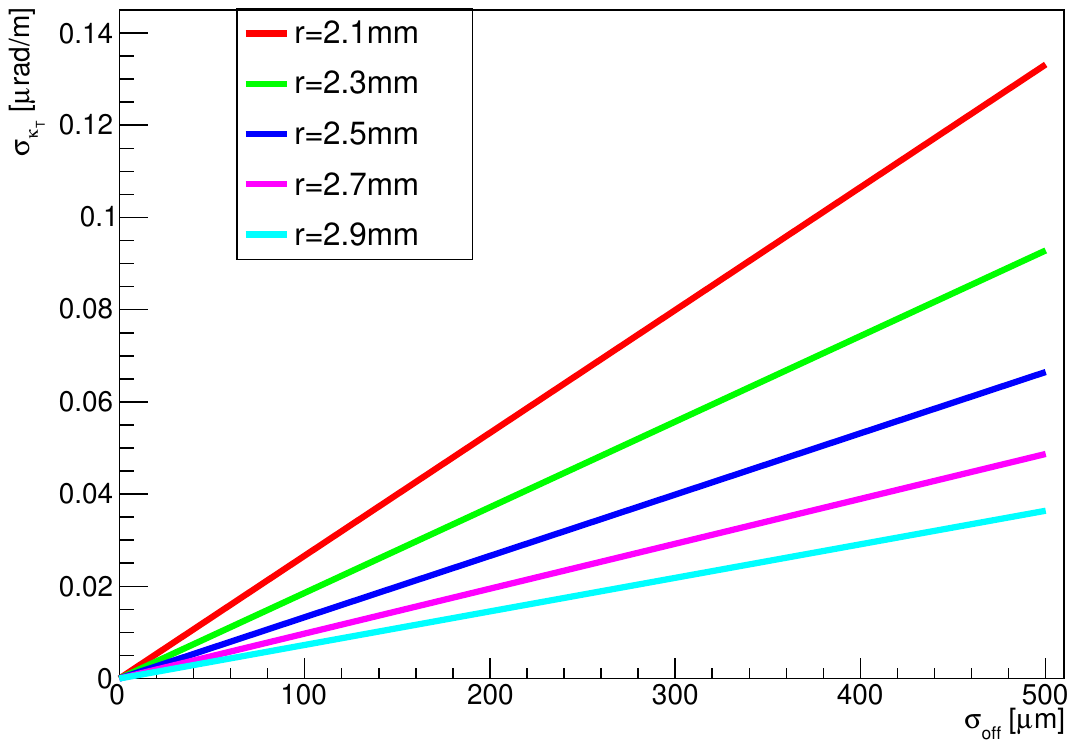}
\caption{Kick angle per unit length uncertainty as a function of the misalignment jitter value, for different vacuum chamber inner radii.}
\label{fig:4}
\end{figure}
to evaluate the kick angle error $\sigma_{\kappa_T}$ as a function of the jitter $\sigma_{\rm off}$ on the transverse offset $\ell_{\rm off}$.
For a given $r$ value in Eq.(\ref{eq:RW2}), $\sigma_{\kappa_T}$ scales linearly with any $\sigma_{\rm off}$ uncertainty.
The $\sigma_{\rm off}$ represents the jitter due to offsets between adjacent undulator modules or between the vacuum chamber and undulator within the same module. The offset variations may occur either within the same FEL shot or between successive shots.

Figure~\ref{fig:4} shows the uncertainty in $\kappa_T$ as a function of the misalignment jitter $\sigma_{\rm off}$ value, for different vacuum chamber inner radii.
As a result, for a VC inner radius of 2.5 mm, an offset jitter of about 300 $\mu$m induces a kick angle error of about 40 nrad$/$m on an as small as 2\% portion of the bunch current distribution.

\section{Trajectory tolerance at undulator entrance}
\label{sec:4}
Electron beam misalignment at injection affects the FEL gain, reducing the maximum power and increasing the saturation length.
The tolerance of the transverse beam position jitter at injection is estimated with the \texttt{Genesis1.3} simulation code,
by evaluating the FEL power starting with an off-axis injection,
along both horizontal and vertical planes.
\begin{figure}[!htb]
\centering
\includegraphics*[width=0.9\columnwidth]{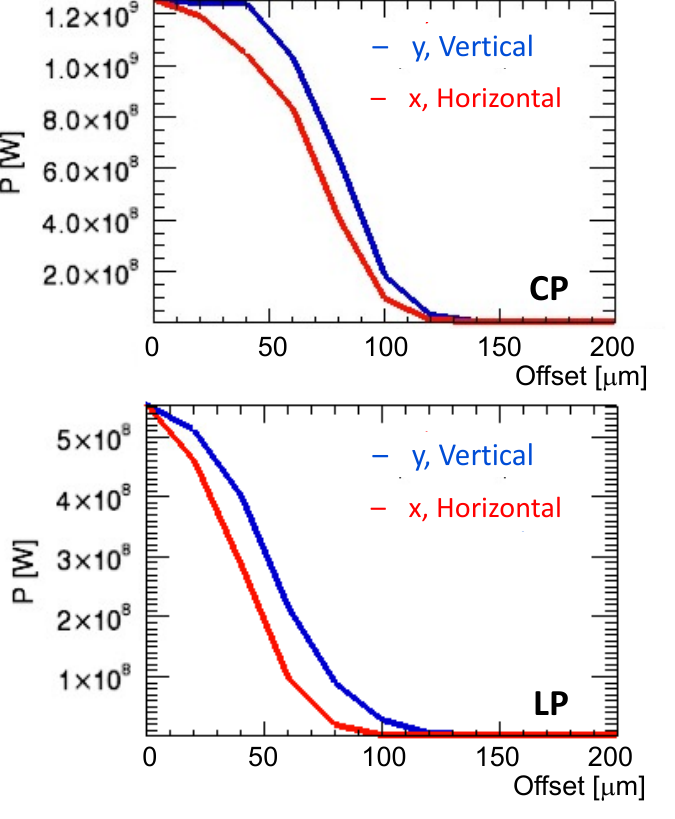}
\caption{FEL power degradation as a function of the transverse injection offset, along either horizontal or vertical plane,
for both circular (top) and linear (down) polarization undulator settings.}
\label{fig:5}
\end{figure}
Figure~\ref{fig:5} shows the FEL power reduction at increasing transverse injection offset, along either horizontal or vertical plane,
for both circular (top panel) and linear (down panel) polarization operations.
\begin{figure}[!htb]
\centering
\includegraphics*[width=0.96\columnwidth]{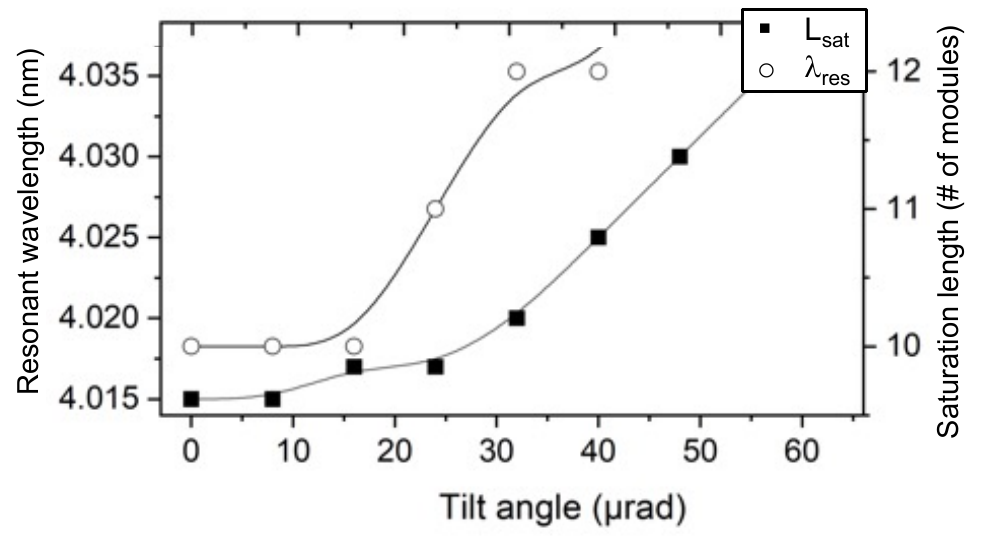}
\caption{Resonant wavelength (full squares) and saturation length (empty circles) in unit of number of undulator modules, as a function of the tilt angle at undulator entrance.}
\label{fig:6}
\end{figure}
An off-axis injection of 50~$\mu$m reduces the FEL power at the undulator exit to a degree that depends on both the undulator settings (CP or LP) and whether the injection occurs in the horizontal or vertical plane. The results in Figure~\ref{fig:5} are summarized as follows:
\begin{description}
\item[CP:] At undulator exit, the FEL power is reduced to 75\% or 90\% of the on-axis peak value when the beam is injected off-axis in the horizontal or vertical plane, respectively.
\item[LP:] At undulator exit, the FEL power is reduced to 34\% or 56\% of the on-axis peak value when the beam is injected off-axis in the horizontal or vertical plane, respectively.
\end{description}
The constraint on the off-axis position at undulator entrance gets even more severe when coupled to a possible tilted injection angle $\theta_{\rm tilt}$.
This latter error source affects also the wavelength spectrum, in addition to reducing the FEL power and to stretching the gain length.
Figure~\ref{fig:6} shows both the resonant wavelength and the saturation length expressed in number of APPLE-X modules, as a function of the tilt angle.
\begin{figure}[!htb]
\centering
\includegraphics*[width=0.95\columnwidth]{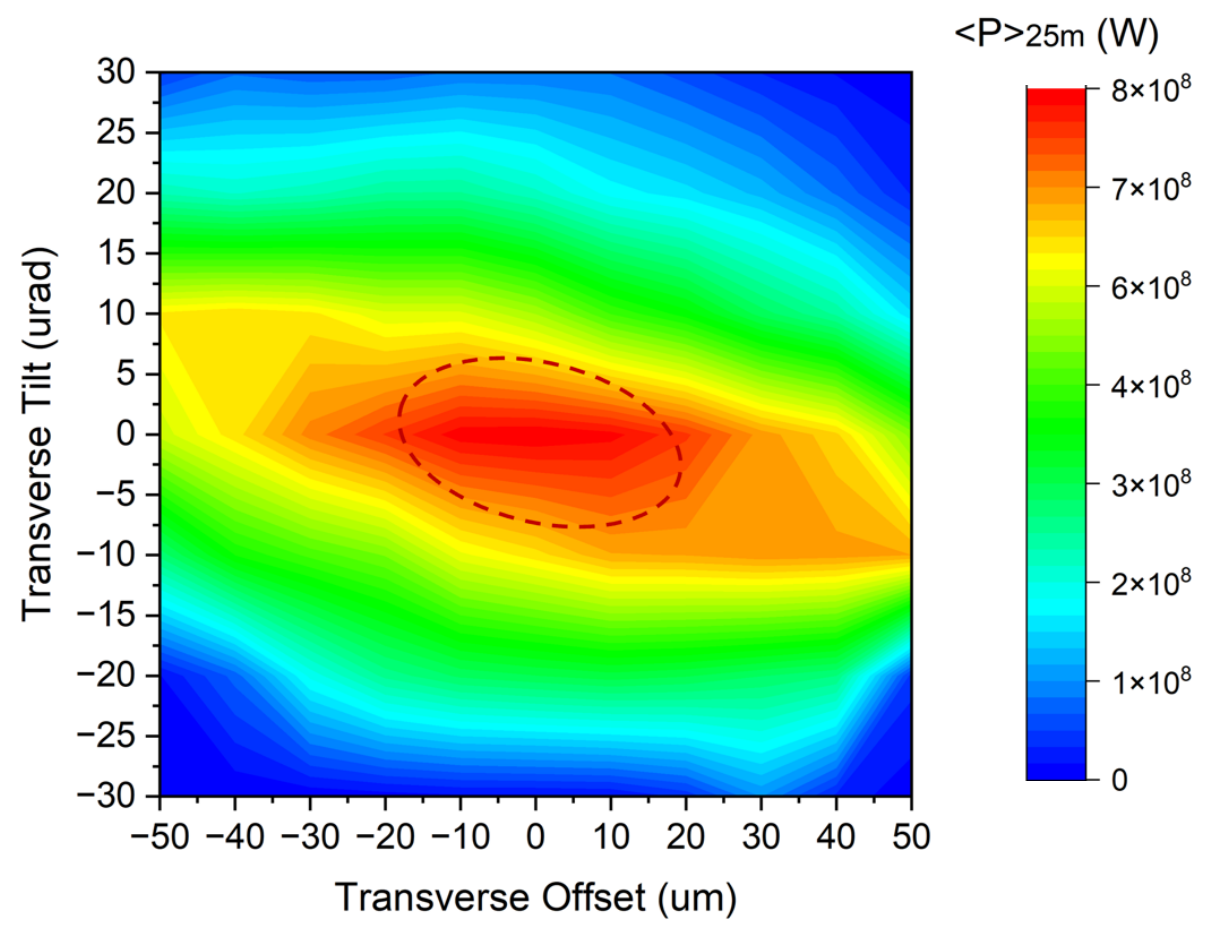}
\caption{Average FEL power (color map) achieved after 25 m, as a function of tilt angle and offset position at undulator entrance.
Exact symmetry in horizontal and vertical planes is assumed for these quantities affecting the injection of the electron beam centroid.}
\label{fig:7}
\end{figure}
Electron beam misalignments due to tilted injections detune the resonant wavelength value and increase
the saturation length. As foreseen in the AQUA design, ten undulator modules demand for $\theta_{\rm tilt}<25\,\mu$rad.
This corresponds to a 0.06\% central wavelength detuning. 
Although this wavelength variation might be accepted, such a tilt angle value affects the FEL power at undulator exit,
further limiting the injection offset position tolerance.
Figure~\ref{fig:7} shows the average FEL power after 25 m of the undulator beamline, as a function of the electron beam tilt angle and offset position at injection.
This correlation between tilt angle and off-axis position results in tighter constraints than previously discussed analyzing each error source independently. To stay with at least 60\% of the ideal FEL peak power, the following tolerance conditions on injection imperfections apply:
$\theta_{\rm tilt}<6\,\mu$rad and $\ell_{\rm off}<25\,\mu$m. The dashed ellipse superimposed on the two-dimensional plot indicates the area of acceptable variations.

\section{Conclusions}
\label{sec:5}
The undulator configuration chosen for the AQUA beamline comprises ten out-of-vacuum APPLE-X modules. This setup provides the capability to generate FEL light tailored for experiments with selectable polarization in the water window spectral range.

The performance of the FEL, designed to operate at a target wavelength of $\lambda = 4$ nm and driven by a 1 GeV electron beam, has been systematically evaluated considering key factors contributing to inhomogeneous broadening, including transverse emittance and energy spread. Operating the beamline with normalized emittance values in the range of 0.6-0.9 mm-mrad, combined with a fractional energy spread between $1.5 \times 10^{-4}$ and $3.5 \times 10^{-4}$, enables achieving a saturation length shorter than 30 meters. This result emphasizes the importance of tight control over beam quality to ensure optimal FEL performance.

Extensive analyses have been conducted on the effects of wakefields, both longitudinal, affecting energy loss, and transverse, impacting the electron trajectory. The results confirm that a vacuum pipe with an inner radius of 2.5 mm represents a reliable choice, balancing the mitigation of wakefield effects with practical design constraints.

Furthermore, 3D time-dependent simulations have been performed to evaluate FEL tolerance to electron beam injection misalignments. These studies provide critical insights into the acceptable ranges for off-axis injection position and tilt angle. The impact of these misalignments has been analyzed in terms of wavelength detuning, saturation length, and power degradation, offering practical guidance for beamline alignment and tuning. The correlation between these parameters reveals that achieving consistent FEL performance demands more stringent constraints on their combined variation than would be required if each parameter were treated independently.

In summary, the results demonstrate that achieving saturation at the target wavelength of 4 nm with an average output power exceeding $10^8$ W is feasible within the capabilities of the layout. 
The robustness of this design is confirmed even when accounting for the combined effects of emittance, energy spread, wakefields and alignment tolerances. These findings consolidate the feasibility of the AQUA beamline to deliver polarized FEL radiation for advanced experiments in the water window.
%
%
%
\section*{Acknowledgements}
This work was partially funded by the European Commission under grant n. 101079773. We wish to thank Mauro Migliorati and Fabio Bosco for providing us some estimates of the energy loss due to resistive wall longitudinal wakefields. 




\end{document}